\documentclass[
,final            
  ]
  {aipproc}

\layoutstyle{8x11double}

\newcommand\sun{\odot}

\begin{document}

\title{Population III Wolf-Rayet Stars in the CAK Regime}


\author{A. J. Onifer}{
  address={Los Alamos National Laboratory, PO Box 1663, MS T087, Los Alamos, NM 87544},
}

\begin{abstract}
Wolf-Rayet (WR) stars near solar metallicity are believed to be driven by radiation
pressure on the UV spectral lines of metal ions.  As the metallicity decreases
so does the line opacity, therefore the mass-loss rate.  However, since the
composition of a WR atmosphere is determined by the burn products of the core,
there is a lower limit on the line opacity -- and therefore the mass-loss rate
-- of a WR star, even in a star with zero initial metallicity.
This presentation is the result of attempt to calculate the mass-loss rate of a
Population III-type WO star using
a modified version of the CAK approximation.  I find that $n_e \ge 10^{13}$ 
cm$^{-3}$ and
$0.5 \leq \Gamma \leq 0.7$ give the most plausible results, with the resulting
mass-loss rate between $2\times10^{-9}$ M$_\sun$ yr$^{-1}$ and $3\times10^{-8}$ 
M$_\sun$ yr$^{-1}$.
\end{abstract}

\maketitle


\section{Introduction}

At low metallicity line-driven Wolf-Rayet (WR) winds exhibit smaller mass-loss
rates than WR stars near solar metallicity.  Since WR winds are significant
contributers of chemical enrichment and energy deposition at all metallicities, 
understanding the physics of wind driving at low metallicity is important.

WR winds have been studied computationally via radiative transfer \cite{hilliermiller98} or 
radiation hydrodynamics simulations \cite{grafenerhamann05}.  
As a complement to these analyses, this
study uses simple analytic models based on a modified version of CAK theory
\cite{cak} to determine the effects of a Population III abundance profile on WR
wind driving.  To keep the results as general as possible,
model-specific parameters such as the velocity law and the location of the
critical point are not specified.

Recently \citet{vinkdekoter05} explored the metallicity dependence of WR mass
loss using a Monte Carlo approach with CAK-type line driving for stars at
metallicity down to $Z / Z_\sun = 10^{-5}$. They found that the mass-loss rate
flattens at low metallicity ($Z / Z_\sun < 10^{-3}$) because
the self-enrichment of N and He in WN stars and C
in WC stars provide the lines needed for wind driving. They found
$\dot{M} \approx 10^{-8}$ M$_\sun yr^{-1}$ for low metallicity WN stars and 
$\dot{M} = 1.4 \times 10^{-7}$ M$_\sun$ yr$^{-1}$ for low metallicity WC stars.
With these results in mind, this presentation provides results for a WO star at
the extreme of zero background metallicity, varying the luminosity (via the
Eddingtom parameter $\Gamma$) and the electron density.

\section{CAK Theory and Modifications}

The first successful analytic models of line-driven winds were developed by \citet{cak} (hereafter CAK) 
to model the winds of OB stars. CAK theory models wind driving via radiation
pressure on a large number of UV lines.  It relies on the Sobolev
approximation \cite{sobolev60}, which assumes a rapidly accelerating wind such
that a
photon's Sobolev length $L_{sob} = v_{th} (dv/dr)^{-1}$, where $v_{th}$ is the
thermal speed and $dv/dr$ is the gradient of the wind velocity, is much smaller
than the mean free path between lines.

In the CAK model the radiation pressure
on the lines is expressed as a ratio of the radiative acceleration of the lines
$g_L$ to the radiative acceleration of the free electrons $g_e$ using the
so-called force multiplier $M(t)$,
\begin{equation}
\label{forcemultdef}
M(t) = \frac{g_L}{g_e}.
\end{equation}
The force multiplier is then parameterized by the relation
\begin{equation}
\label{forcemultparam}
M(t) = k \: t^{-\alpha},
\end{equation}
where $k$ and $\alpha$ are determined from the line strength distribution, and
$\alpha$ in particular is a number between 0 and 1 that is related to the
fraction of optically thick lines, thus $\alpha = 1$ means all lines are
optically thick over $L_{sob}$.

The details of
the model used in this analysis, which include the effects of frequency
redistribution due to line branching and bound-free opacity, can be found in 
\citet{onifergayley06} (hereafter OG) and references therein.  The following are some highlights.

In CAK theory the 
momentum equation at the critical point in
units of the effective gravity can be written (see OG, Sec. 2 and Eq. 27),
\begin{equation}
\label{cakmomeq}
1 + y_c = \frac{\Gamma}{1 - \Gamma} \; M(t) \; F_{NID} \; F_{red},
\end{equation}
where the first term on the left-hand side represents the effective gravity, $\Gamma$ is the Eddington parameter, $F_{NID}$ is a correction for multiple
scattering \cite{gayleyetal95}, $F_{red}$ is a correction for frequency
redistribution due to line branching and bound-free opacity, and $y$
is the inertia scaled to the effective gravity,
\begin{equation}
\label{ydef}
y_c = \frac{r_c^2 \; v_c \; (dv/dr)_{r_c}}{G M_* (1 - \Gamma)},
\end{equation}
where $r_c$ and $v_c$ are the radius and velocity at the critical point and
$M_*$ is the stellar mass.

\section{Model Setup}

One cannot hope to model analytically all the complexity 
that a radiative tranfer or radiation hydrodynamics code is
capable of modeling, so some simplifying assumptions must be made.
The wind is assumed to be spherically symmetric, smooth on scales
larger than $L_{sob}$, and time steady, and the star is assumed to be
nonrotating.

At low $Z$ a WR star is more likely to evolve to the WO phase
than at high $Z$ because the small amount of mass lost in the WN
phase allows more production of O via $\alpha$-particle capture of C before the
overlying envelope is stripped.  The WO phase can be defined
by the relation $(N_C + N_O) / N_{He} > 1.0$, where $N_i$ is surface number fraction \cite{barlowhummer82}.
The surface abundances are based on the SMC WO star Sand 1 studied by
\citet{kingsburghetal95}.
The mass fractions of the  
included elements are $Y = 0.21$, $X_C = 0.51$, $X_O = 0.25$, $X_{Ne} = 
0.02$.  The Ne is entirely $^{22}$Ne, which is a product of 
the He-burning process. All ion states of each 
element are included.  The Sobolev opacities are calculated using oscillator
strengths from the Opacity Project \cite{badnellseaton05}.  Details of the 
opacity calculation can be found in OG.

The mass of the star is 5 $M_\sun$, also based on Sand 1. This is may be a bit
small even for a highly-evolved Population III WR
star due to the lower mass-loss rates of such stars, so the mass-loss rates should be
thought of as lower limits.
The terminal speed $v_{\inf}$ is set such that $v_{\inf} - v_{c}
= 4700$ km s$^{-1}$. 
\citet{kingsburghetal95} report $v_{\inf} = 4200$ km s$^{-1}$, which gives $v_c =
500$ km s$^{-1}$. This reflects the large terminal speeds of WO stars. 
$\dot{M}$, though ultimately has a small dependence on $v_{\inf}$, since $\dot{M}$ is
set by conditions near the surface at the critical point and $v_{\inf}$ is set by the global
wind conditions.
The temperature at the critical point $T_c =
1.3 \times 10^5$ K.  Two sets of
models are run.  One set assumes complete frequency redistribution (CRD).
The other relaxes the CRD assumption by accounting for
for changes in the wind
ionization structure due to the changing wind temperature over a 
thermalization length, or length beyond which frequency redistribution
affects the spectral flux profile \cite{lucyabbott93}.  In this model the
thermalization length is such that the final temperature is $T =
4.0\times10^{4}$ K.  The electron density $n_e$ is set to either $10^{12}$
cm$^{-3},
10^{13}$ cm$^{-3}$ or $10^{14}$ cm$^{-3}$.  Setting a constant $n_e$ implies that the
velocity curve and/or the stellar radius is different for each model, though
neither as specified explicitly, as this study is concerned more with modeling a
representative star, rather than a grid of specific stars.

\section{Results}

\begin{figure}
  \protect\label{tausob1}
  \includegraphics[height=.20\textheight]{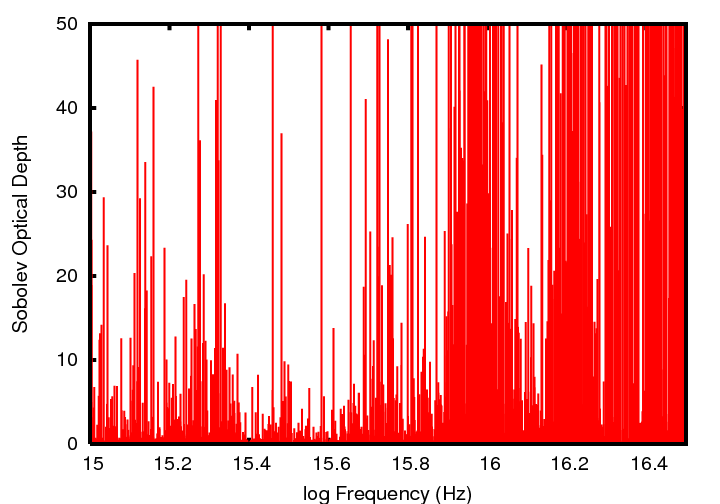}
  \caption{Sobolev optical depth as a function of frequency at solar
metallicity.}
\end{figure}

\begin{figure}
  \protect\label{tausobz0}
  \includegraphics[height=.20\textheight]{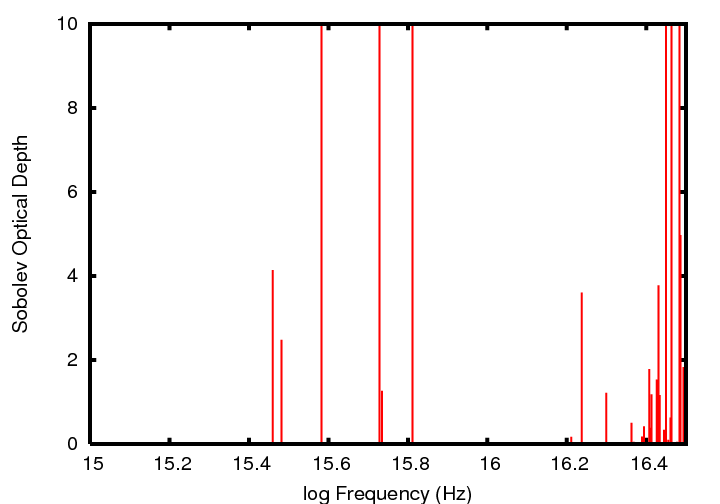}
  \caption{The same as figure \ref{tausob1}, but at Z = 0 and CRD.}
\end{figure}

Not surprisingly, the lack of metals, especially of iron group ions, causes a
large drop in the line opacity.  Figure \ref{tausob1} shows the Sobolev optical depth
spectrum for a solar metallicity star with stellar parameters similar to those
explored in this paper.  Figure \ref{tausobz0} shows the Sobolev optical depths
for the $Z = 0$ case in CRD.  Clearly there is much less opacity to drive the wind.

The mass loss rates and wind parameters are shown in table \ref{tableg5}.   Only
models that produced physical results (e.g., $F_{red} < 1$) are shown.  None of
the CRD models were able to produce a wind.  Thus for the given parameter
ranges a CAK-type wind requires a finite thermalization length.  As the wind density
drops, $M(t)$ rises, meaning when the continuum is weaker, the
lines take on a larger fraction of the momentum driving.  The models
for which $10 < M(t) < 50$ are most similar to Population I WR stars, and are
probably the most plausible for Population III WR stars.  Of these, the
mass-loss rate is in the range $2\times10^{-9}$ M$_\sun$ yr$^{-1} < \dot{M}
< 3\times10^{-8}$ M$_\sun$ yr$^{-1}$.  Compared to Vink and de Koter's low-Z mass-loss rate $\dot{M}
\approx 10^{-7}$ M$_\sun$ yr$^{1}$, my rates are small, though the results are
not directly comparable due to the difference in stellar mass and 
the WO-like abundances of models in this presentation.


\begin{table}
\begin{tabular}{rrrrrl}
\hline
    \tablehead{1}{c}{b}{$\mathbf{n_e}$ (cm$\mathbf{^{-3}}$)}
  & \tablehead{1}{c}{b}{${\Gamma}$} 
  & \tablehead{1}{c}{b}{$\mathbf{\tau_e}$} 
  & \tablehead{1}{c}{b}{{\it\bf t}}   
  & \tablehead{1}{c}{b}{{\it\bf M(t)}} 
  & \tablehead{1}{c}{b}{$\mathbf{\dot{M}}$ (M$\mathbf{_\sun}$ yr$\mathbf{^{-1})}$}\\
\hline
$1\times10^{14}$ & 0.3 & 5  & $3.6\times10^{-5}$ & 96.3 & $7.3\times10^{-9}$\\
$1\times10^{14}$ & 0.3 & 10 & $4.1\times10^{-5}$ & 91.5 & $8.6\times10^{-9}$\\
$1\times10^{14}$ & 0.5 & 5  & $1.0\times10^{-4}$ & 6.7  & $1.6\times10^{-8}$\\
$1\times10^{14}$ & 0.5 & 10 & $1.1\times10^{-4}$ & 12.0 & $1.9\times10^{-8}$\\
$1\times10^{14}$ & 0.7 & 2  & $5.0\times10^{-4}$ & 15.0 & $1.9\times10^{-8}$\\
$1\times10^{14}$ & 0.7 & 5  & $7.1\times10^{-4}$ & 13.5 & $2.5\times10^{-8}$\\
$1\times10^{14}$ & 0.7 & 10 & $8.9\times10^{-4}$ & 10.6 & $2.8\times10^{-8}$\\
\hline   
$1\times10^{13}$ & 0.3 & 5  & $1.2\times10^{-5}$ & 134  & $8.6\times10^{-10}$\\
$1\times10^{13}$ & 0.3 & 10 & $1.7\times10^{-5}$ & 98.9 & $1.0\times10^{-9}$\\
$1\times10^{13}$ & 0.5 & 2  & $5.8\times10^{-5}$ & 49.7 & $2.2\times10^{-9}$\\
$1\times10^{13}$ & 0.5 & 5  & $7.5\times10^{-5}$ & 31.5 & $3.5\times10^{-9}$\\
$1\times10^{13}$ & 0.5 & 10 & $8.6\times10^{-5}$ & 26.4 & $4.5\times10^{-9}$\\
$1\times10^{13}$ & 0.7 & 2  & $1.5\times10^{-4}$ & 17.2 & $6.5\times10^{-9}$\\
$1\times10^{13}$ & 0.7 & 5  & $1.7\times10^{-4}$ & 15.8 & $9.5\times10^{-9}$\\
$1\times10^{13}$ & 0.7 & 10 & $1.8\times10^{-4}$ & 15.1 & $1.2\times10^{-8}$\\
\hline   
$1\times10^{12}$ & 0.3 & 2  & $2.8\times10^{-6}$ & 377  & $2.8\times10^{-10}$\\
$1\times10^{12}$ & 0.3 & 5  & $3.1\times10^{-6}$ & 394  & $3.9\times10^{-10}$\\
$1\times10^{12}$ & 0.3 & 10 & $3.3\times10^{-6}$ & 326  & $4.6\times10^{-10}$\\
$1\times10^{12}$ & 0.5 & 2  & $5.1\times10^{-6}$ & 222  & $7.2\times10^{-10}$\\
$1\times10^{12}$ & 0.5 & 10 & $5.8\times10^{-6}$ & 225  & $1.1\times10^{-9}$\\
\hline
\end{tabular}
\caption{Results for model calculations}
\label{tableg5}
\end{table}

\section{Conclusions}

A small parameter study of an analytic,
modified CAK-type model of Population III WO-type stellar winds has been performed.
The most plausible results were found for $\Gamma = 0.5$ and $\Gamma = 0.7$ and
$10^{13}$ cm$^{-3} < n_e < 10^{14}$ cm$^{-3}$, with mass-loss rates in the range
$2\times10^{-9}$ M$_\sun$ yr$^{-1} < \dot{M} < 3\times10^{-8}$ M$_\sun$ yr$^{-1}$.

A similar analysis could be done for WN stars, a phase through which all WR
stars pass.  WN stars will have an even lower mass-loss rate, as the only line
driving would be provided by He and a relatively small amount of N. These
results will be used in a study of mass-loss as a function of small but nonzero
metallicity.


\begin{theacknowledgments}
The author would like to thank Alexander Heger for helpful discussions.  This work was performed 
under the auspices of the U. S. Department of Energy by the Los Alamos National Security (LANS), LLC under contract No. DE-AC52-06NA25396.
\end{theacknowledgments}



\bibliographystyle{aipproc}   

\bibliography{onifer-astroph}

\IfFileExists{\jobname.bbl}{}
 {\typeout{}
  \typeout{******************************************}
  \typeout{** Please run "bibtex \jobname" to optain}
  \typeout{** the bibliography and then re-run LaTeX}
  \typeout{** twice to fix the references!}
  \typeout{******************************************}
  \typeout{}
 }

\end{document}